\documentclass{article}
\usepackage[noadjust]{cite}

\usepackage[T1]{fontenc}
\usepackage{lineno} 
\usepackage{comment}

\usepackage{amsmath,amssymb}
\usepackage{color,soul}
\usepackage{calrsfs}
\usepackage[cmintegrals]{newtxmath}
\usepackage{appendix}

\usepackage{spconf,amsmath,graphicx}
\usepackage{enumitem}
\usepackage{comment}
\usepackage{amsmath,amssymb,epsfig}
\usepackage{bm}
\usepackage{verbatim}
\usepackage{ieeetrantools}
\usepackage{setspace}
\usepackage{subfig}
\usepackage[x11names]{xcolor}

\usepackage[belowskip=2pt,aboveskip=10pt]{caption}

\DeclareMathAlphabet{\pazocal}{OMS}{zplm}{m}{n}
\SetMathAlphabet\pazocal{bold}{OMS}{zplm}{bx}{n}

\DeclareMathOperator*{\argminA}{argmin}
\newcommand{\bbR}{\ensuremath{\mathbb{R}}}
\newcommand{\bbE}{\ensuremath{\mathbb{E}}}
\newcommand{\bbN}{\ensuremath{\mathbb{N}}}
\newcommand{\bbC}{\ensuremath{\mathbb{C}}}
\newcommand{\cW}{\ensuremath{\pazocal{W}}}

\newcommand{\ba}{\ensuremath{\mathbf{a}}}

\newcommand{\bh}{\ensuremath{\mathbf{h}}}

\newcommand{\bq}{\ensuremath{\mathbf{q}}}

\newcommand{\be}{\ensuremath{\mathbf{e}}}

\newcommand{\bd}{\ensuremath{\mathbf{d}}}

\newcommand{\balpha}{\ensuremath{\boldsymbol{\alpha}}}

\newcommand{\bphi}{\ensuremath{\boldsymbol{\phi}}}

\newcommand{\btau}{\ensuremath{\boldsymbol{\tau}}}
\newcommand{\bsigma}{\ensuremath{\boldsymbol{\sigma}}}
\newcommand{\bgama}{\ensuremath{\boldsymbol{\gamma}}}

\newcommand{\bTheta}{\ensuremath{\mathbf{\Theta}}}
\newcommand{\bLambda}{\ensuremath{\mathbf{\Lambda}}}
\newcommand{\bW}{\ensuremath{\mathbf{W}}}

\newcommand{\bM}{\ensuremath{\mathbf{M}}}
\newcommand{\bP}{\ensuremath{\mathbf{P}}}
\newcommand{\bH}{\ensuremath{\mathbf{H}}}

\newcommand{\bE}{\ensuremath{\mathbf{E}}}
\newcommand{\bA}{\ensuremath{\mathbf{A}}}
\newcommand{\bR}{\ensuremath{\mathbf{R}}}

\newcommand{\bD}{\ensuremath{\mathbf{D}}}

\newcommand{\bI}{\ensuremath{\mathbf{I}}}
\newcommand{\bJ}{\ensuremath{\mathbf{J}}}

\newcommand{\bU}{\ensuremath{\mathbf{U}}}
\newcommand{\bu}{\ensuremath{\mathbf{u}}}
\newcommand{\bzero}{\ensuremath{\mathbf{0}}}

\newif\ifproofread

    \title{Analysis of multipath channel delay estimation using subspace fitting\footnotemark}

    \name{Tarik Kazaz\IEEEauthorrefmark{1}, Jac Romme\IEEEauthorrefmark{2}, Gerard J. M. Janssen\IEEEauthorrefmark{1} and Alle-Jan van der Veen\IEEEauthorrefmark{1}
    \thanks{This research was supported in part by NWO-STW under 
    	contract 13970 (``SuperGPS'').}	}
    \address{\IEEEauthorrefmark{1}Faculty of EEMCS, Delft University of Technology, Delft, The Netherlands \\ 
             \IEEEauthorrefmark{2}Holst Centre - IMEC-NL, Eindhoven, The Netherlands}

\begin{document}
\proofreadfalse
\maketitle


\begin{abstract}
	\noindent 
    The presence of rich scattering in indoor and urban radio propagation scenarios may cause a high arrival density of multipath components (MPCs). Often the MPCs arrive in clusters at the receiver, where MPCs within one cluster have similar angles and delays. The MPCs arriving within a single cluster are typically unresolvable in the delay domain. In this paper, we analyze the effects of unresolved MPCs on the bias of the delay estimation with a multiband subspace fitting algorithm.  We treat the unresolved MPCs as a model error that results in perturbed subspace estimation. Starting from the first-order approximation of the perturbations, we derive the bias of the delay estimate of the line-of-sight (LOS) component. We show that it depends on the power and relative delay of the unresolved  MPCs in the first cluster compared to the LOS component. Numerical experiments are included to show that the derived expression for the bias well describes the effects of unresolved MPCs on the delay estimation.
\end{abstract}


\section{Introduction}
\begin{figure*}[t]
    	\centering
    	 \includegraphics[trim=1 1 0 1,clip, width=15cm]{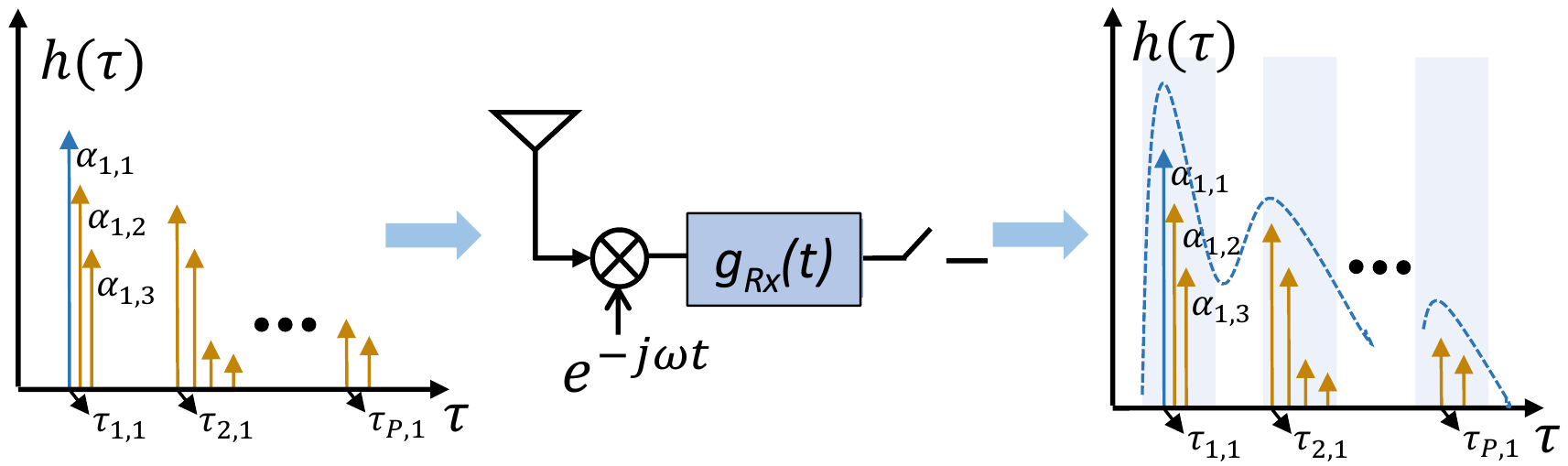}%
    	 \caption{The multipath channel estimation and clustering of dense MPCs due to the limited bandwidth of transceiver chains.}
    	\label{fig:ilu}
    \end{figure*}

    \noindent The delay estimation for time-of-arrival (TOA) localization starts with channel probing and estimation of multipath components (MPCs) parameters. Since the radio transceivers used for channel probing have limited bandwidth $B$, it is challenging to achieve high-resolution delay estimation from bandlimited channel measurements (cf. Fig. \ref{fig:ilu}). When MPCs are well separated in the delay domain, a unique solution for estimates of delay and corresponding complex gain parameters of MPCs can be obtained. However, in practical urban and indoor propagation scenarios, MPCs typically exhibit diffuse scattering. This results in a partial overlap of MPCs in angle and delay domains and the arrival of clusters of MPCs at the receiver, where MPCs within one cluster have similar angles and delays. 
    
   Typically, the clustering of MPCs is modeled using the extended Saleh-Valenzuela model \cite{saleh1987statistical}. This model defines the channel impulse response (CIR) as a sparse sequence of clusters of MPCs. With this modeling assumption, the delay estimation of MPCs becomes a problem of parametric spectral inference from observed measurements. Then, algorithms for delay estimation based on (i) subspace estimation  \cite{van1998joint, kazaz2018joint, kazaz2019multiresolution}, (ii) finite rate of innovation \cite{vetterli2002sampling, gedalyahu2010time, gedalyahu2011multichannel}, or (iii) compressed sampling \cite{zhang2009compressed, kazaz2019jointcal}, can be used to increase the temporal resolution of delay estimation. However, when intra-cluster MPCs have much smaller separation in the delay domain compared to $\frac{1}{B}$, it is impossible to resolve these MPCs. As a result, multiple closely arriving MPCs are classified as a single MPC, which leads to a biased delay estimation. 
    
    The performance analysis and statistical efficiency of the subspace-based methods have been discussed in \cite{viberg1991detection, viberg1991sensor, stoica1989music, kangas1994finite}. In \cite{astely1999effects, lin2007further, liu2008first}, the effects of local scattering on the direction-of-arrival (DOA) estimation using MUSIC and ESPRIT algorithms have been analyzed. Performance analysis of multidimensional subspace-based algorithms in the presence of model errors is discussed in \cite{swindlehurst1993performance}. However, the effects of unresolved MPCs on delay estimation using subspace fitting methods have not been studied before.

    In this work, we analyze the effects of dense multipath components on the bias of the delay estimation using the multiband subspace fitting method \cite{kazaz2019asilomar}. For TOA localization, only delay estimation of the LOS component is of interest, while the other unresolved MPCs within the first cluster can be considered as interference. We treat this as a model error, which results in the perturbed subspace estimation, leading to a bias in delay estimation. We assume that the delay spread is small and use the first-order approximation of perturbation to derive the bias of the delay estimates. It can be seen that the bias in the delay estimate of the LOS component depends on the power and relative delay of unresolved MPCs compared to the LOS component. 
    
    Numerical simulations are conducted to verify the analytical results. It is shown that derived expression for the bias well describes the effects of unresolved MPCs on delay estimation.
    
	\section{Data Model}
	\noindent Typically, the clustered multipath channel is modeled using the extended Saleh-Valenzuela model \cite{saleh1987statistical} as 
	\begin{equation}
        \label{eq:chan_imp}
        h(t) = \sum_{p=1}^{P}\sum_{k=1}^{K_p} \alpha_{p,k} \delta(t-\tau_{p,k})\,, 
    \end{equation}
    \noindent where $P$ is the number of clusters, $K_p$ is the number of MPCs in the $p$th cluster, $K = \sum_{p=1}^{P}K_p$ is the total number of MPCs in the channel, $\alpha_{p,k} \in \bbC$ and $\tau_{p,k} \in \bbR_+$ are the unknown gain and time-delay of $k$th MPC in the $p$th cluster (cf. Fig. \ref{fig:ilu}), respectively. The corresponding frequency response of the channel is given by 
    \begin{equation}
        \label{eq:chan_freq}
        H(\omega) = \sum_{p=1}^{P}\sum_{k=1}^{K_p} \alpha_{p,k} e^{-j\omega\tau_{p,k}}.
    \end{equation}
    
    Let us assume that the channel is sampled in the frequency domain using wideband OFDM probing signals with $N$ sub-carriers transmitted over $i = 0, \dots, L-1$ separate frequency bands \cite{kazaz2019asilomar}. The probed frequency bands are $\cW_i = [\omega_i - \frac{B}{2}, \omega_i + \frac{B}{2}]$, where $B$ is the bandwidth, and $\omega_i$ is the central angular frequency of the $i$th band. We consider that the bands $\left\{\cW_i\right\}_{i=0}^{L-1}$ are lying on the discrete frequency grid $\omega_i = \omega_0 + n_i\omega_{\rm s}$, where $n_i \in \bbN$, $\omega_{0}$ denotes the lowest frequency used during channel probing, and $\omega_s$ is the sub-carrier spacing. The receiver estimates the channel frequency responses at the $N$ subcarrier frequencies  $\bh_i \in \bbC^{N}$, in each of the bands $i=0,\dots,L-1$. The estimated channel frequency responses are collected in the multiband channel vector $\bh = [\bh_0^T,\dots,\bh_{L-1}^T]^T \in \bbC^{NL}$, which satisfies the model \cite{kazaz2019asilomar}
    \begin{equation}
        \label{eq:multiband}
            \bh = \bA(\bphi)\balpha + \bq := 
     \begin{bmatrix} 
          \bM \\
          \bM\bTheta_1\\
          \vdots\\
          \bM\bTheta_{L-1}\\
      \end{bmatrix}\balpha + 
      \begin{bmatrix}
          \bq_{0} \\ \bq_{1} \\ \vdots \\ \bq_{L-1}
      \end{bmatrix} \,,
    \end{equation}
    where  $\bA(\bphi) = [\ba(\phi_{1,1}),\dots,\ba(\phi_{P,K})] \in \bbC^{NL \times K}$, $\bM \in \bbC^{N\times K}$ is a Vandermonde matrix
    \begin{equation}
	\label{eq:f_matrix}
	\bM = 
	\begin{bmatrix}
	1 & 1 & \cdots & 1 \\
	\Phi_{1,1} & \Phi_{1,2} & \cdots & \Phi_{P,K_P}\\
	\vdots & \vdots & \ddots & \vdots \\ 
	\Phi_{1,1}^{{N-1}} & \Phi_{1,2}^{N-1} & \cdots & \Phi_{P,K_P}^{N-1}
	\end{bmatrix}
	\,,\quad
	\end{equation}
    	$ \Phi_{p,k} = e^{-j\phi_{p,k}}$, and $\phi_{p,k} = \omega_{s} \tau_{p,k}$. Likewise, $\bTheta_i = \text{diag}([\Phi_{1,1}^{n_i},\dots,\Phi_{P,K}^{n_i}]) \in \bbC^{K\times K}$  are band dependent phase shifts, $\balpha = [\alpha_{1,1},\dots,\alpha_{P,K_P}]^T \in \bbC^{K}$ are gains of the MPCs, and $\bq$ is a zero-mean Gaussian noise vector.  Since $\bA(\bphi)$ has a multiple shift-invariance structure, and (\ref{eq:multiband}) resembles the data model of Multiple Invariance ESPRIT \cite{miesprit1992}, $\bphi$ can be estimated using subspace fitting methods. From $\bphi$, $\btau = [\tau_{1,1},\dots,\tau_{P,K_P}]^T \in \bbR_+^K$ immediately follow. 
	
	We assume that MPCs within a single cluster $p$ are having similar delays, i.e., $\tau_{p,k} \approx \tau_{p,1} + \Delta \tau_{p,k}$, where $\Delta \tau_{p,k} = \tau_{p,k} - \tau_{p,1}$, and it is small compared to $\frac{1}{B}$ $\forall k$. Further, we assume that the clusters of MPCs are sufficiently separated in time, such that they can be successfully resolved, i.e., $(\tau_{p+1,1}-\tau_{p,K_p}) \geq \frac{1}{B}$, $p=1,\dots,P$. Using these assumptions, we can approximate the steering vectors corresponding to the same cluster with their first-order Taylor series expansion as
	\begin{equation}
	    \nonumber
        \label{eq:steering_vec_aprx}
            \ba(\phi_{p,k}) \approx \ba(\phi_{p,1}) + \Delta \phi_{p,k} \bd(\phi_{p,1})\,,
    \end{equation}
    \noindent where $\Delta \phi_{p,k} = \phi_{p,k} - \phi_{p,1}$, and $\bd(\phi_{p,1}) = \frac{\partial \ba(\phi)}{\partial \phi}\vert_{\phi = \phi_{p,1}}$. Then, (\ref{eq:multiband}) can be approximated as
	\begin{equation}
        \label{eq:multiband_aprx}
            \bh \approx [\bA(\tilde{\bphi}) + \bD \text{diag}(\bgama)] \tilde{\balpha} + \bq \,,
    \end{equation}
    \noindent where $\bA(\tilde{\bphi}) = [\ba(\phi_{1,1}), \ba(\phi_{2,1}), \dots,\ba(\phi_{P,1})] \in \bbC^{NL \times P}$ and $\bD = [\bd(\phi_{1,1}), \dots, \bd(\phi_{P,1})] \in \bbC^{NL \times P}$. The elements of vectors $\bgama = [\gamma_1, \dots, \gamma_P]^T$ and $\tilde{\balpha} = [\alpha_1, \dots, \alpha_P]^T$ are given by
    \begin{equation}
    \nonumber
        \label{eq:der_steering2}
        \alpha_p = \sum_{k=1}^{K_p} \alpha_{p,k}, \quad \gamma_p = \frac{\sum_{k=2}^{K_p} \alpha_{p,k} \Delta \phi_{p,k}}{\alpha_p}\,.
    \end{equation}
    
    \section{Performance analysis}
    \noindent Our objective is to analyze the performance of the algorithm for delay estimation as proposed in \cite{kazaz2019asilomar}. We begin by assuming that the main cause for the error is a perturbation of the estimated signal and noise subspaces introduced by unresolved MPCs. Therefore, we neglect the error introduced by finite sampling effects and establish a relation between perturbation of the covariance matrix and the perturbations of the estimated signal and noise subspaces. These results are then used to find the bias of the delay estimator. 
    
    \subsection{First-order Subspace Perturbations}
    The estimation of delays using the multiband estimation algorithm \cite{kazaz2019asilomar} starts with the estimation of the signal subspace from block matrix $\bH \in \bbC^{LM \times Q}$, where $M$ is the design parameter and $Q=N-M+1$. The design parameter must be selected such that $M > K$ and $Q \geq K$ (cf. \cite{kazaz2019asilomar} for details). The estimated covariance matrix of $\bH$ is ${\bR = \frac{1}{Q}\bH \bH^H}$ and satisfies the model 
    \begin{equation}
        \label{eq:cov_matrix}
        \bR = \bA(\tilde{\bphi}) \bR_{\tilde{\alpha}} \bA^H(\tilde{\bphi}) + \bE + \sigma_n^2 \bI\,,
    \end{equation}
    \noindent where $\bR_{\tilde{\alpha}} = \bbE\{\tilde{\balpha} \tilde{\balpha}^H \}$, $\bE=\bA(\tilde{\bphi})\text{diag}(\be)\bD^H + \bD\text{diag}(\be)\bA^H(\tilde{\bphi})$ is the perturbation matrix, and $\sigma_n^2$ is the noise power. Here, we have ignored the second order term of $\Delta \phi_{p,k}$, as it is assumed to be small. Under the assumption of wide-sense stationary uncorrelated scattering (WSSUS) it can be shown that $\bR_{\tilde{\alpha}} = \text{diag}(\bsigma_{\tilde{\alpha}})$, $\bsigma_{\tilde{\alpha}}=[\sigma_1^2, \dots, \sigma_P^2]^T$, where $\sigma_p^2 = \sum_{k=1}^{K_p}\mid\alpha_{p,k}\mid^2$ $\forall p$.
    Likewise, $\be = [e_1, \dots, e_P]^T$, where $e_p = \sum_{k=2}^{K_p} \mid\alpha_{p,k}\mid^2 \Delta \phi_{p,k}$. The $P$-dimensional orthonormal basis for the signal subspace of the estimated covariance matrix $\bR$ when there are no perturbations can be estimated by its eigenvalue decomposition as
    \begin{equation}
        \label{eq:eig_decom}
        \bR = \bU_s\bLambda_s \bU_s^H + \sigma_n^2\bU_n\bU_n^H\,,
    \end{equation}
    \noindent where $\bLambda_s = \text{diag}([\lambda_1,\dots,\lambda_P])$ are the eigenvalues associated to the eigenvectors in $\bU_s = [\bu_{s,1},\dots,\bu_{s,P}]$ and the columns of $\bU_n = [\bu_{n,1},\dots,\bu_{n,LM-P}]$ span the noise subspace of $\bR$. However, when unresolved MPCs introduce perturbations $\bE$, the estimated signal subspace is also perturbed, and we defined it as $\bU_{E} = \bU_s + \bU_e$, where $\bU_e$ is perturbation matrix. The first-order Taylor series expansion of the columns of the perturbed subspace is given in \cite{wilkinson1965algebraic, liu2008first} as
    \begin{equation}
        \label{eq:sub_preturb}
        \bu_{E,i} \approx \bu_{s,i} + \bu_{e,i}\,,
    \end{equation} 
    \noindent where
    \begin{equation}
        \nonumber
        \label{eq:sub_preturb2}
        \bu_{e,i} = \sum_{p=1,i \neq p}^{P} \rho_{i,p} \bu_{s,p} + \sum_{m=1}^{LM-P} \beta_{i,m} \bu_{n,m}\,,
    \end{equation} 
    \begin{equation}
        \nonumber
        \label{eq:sub_preturb3}
        \rho_{i,p} = \dfrac{\bu^H_{s,p}\bE\bu_{s,i}}{\lambda_i - \lambda_p} \quad \text{and}\,, \quad \beta_{i,m} = \dfrac{\bu^H_{n,m}\bE\bu_{s,i}}{\lambda_i}\,.
    \end{equation}

    \subsection{The Multiband Delay Estimation Algorithm}
    
    \noindent The multiband delay estimation algorithm \cite{kazaz2019asilomar} estimates $\bphi$ by solving the weighted subspace fitting problem 
    \noindent 
    \begin{equation}
        \label{eq:subspace_fitting}
        \hat{\bphi} = \argminA_{\bphi} \{\bJ(\bphi)\} = \argminA_{\bphi} \text{tr} \{\bP_A^{\perp}(\bphi)\bU_{E}\bW\bU^H_{E}\}\,,
    \end{equation} 
    \noindent where $\bP_{\bA}^{\perp}(\bphi) = \bI - \bP_{\bA}(\bphi)$, $\bP_{\bA}(\bphi)$ is the projection matrix onto the column space of $\bA(\bphi)$, and $\bW$ is the weighting matrix. Since, $\hat{\bphi}$ minimizes $\bJ(\bphi)$, a first-order Taylor series expansion of $\bJ(\bphi)$ around the true value $\bphi_0$ is given by
    \begin{equation}
        \label{eq:subspace_apprx}
        \bzero = \bJ'(\bphi_0) + \overline{\bJ}''(\bphi_0) (\hat{\bphi}-\bphi_0)\,,
    \end{equation} 
    \noindent where $\bJ'(\bphi_0) = \frac{\partial \bJ(\bphi)}{\partial \bphi}\vert_{\bphi = \bphi_0}$ is the gradient of $J(\bphi)$, $\overline{\bJ}''(\bphi_0) = \lim_{\Delta\bphi \to \bzero}\bJ''(\bphi)\vert_{\bphi = \bphi_0}$, and $\bJ''(\bphi)$ is the Hessian of $\bJ(\bphi)$. The gradient and Hessian of $\bJ(\bphi)$ have been computed in \cite{viberg1991detection, viberg1991sensor}, and they are given by
    \begin{equation}
        \label{eq:gradient_hessian}
        \begin{split}
        & \bJ'(\bphi) = 2\text{Re}\left [\text{diag}(\bA^{\dagger}\bU_E\bW\bU^H_E\bP_A^{\perp}\bD)\right]\,, 
        \\
        \overline{\bJ}''(\bphi) & = - 2\text{Re} \left\{\left[\bD^H\bP_A^{\perp}\bD\right] \odot\left[\bA^{\dagger}\bU_E\bW\bU_E^H(\bA^{\dagger})^H\right]^T\right\}\,,
        \end{split}
    \end{equation} 
    \noindent where $(\cdot)^{\dagger}$ denotes the pseudoinverse of a matrix and $\odot$ is the Kathri-Rao product. Now, from (\ref{eq:subspace_apprx}), the expression for the first-order error, i.e. the bias, is 
    \begin{equation}
        \label{eq:error}
        BIAS(\hat{\bphi}) := \mid \hat{\bphi}-\bphi_0 \mid \approx \left| \overline{\bJ}''^{-1}(\bphi_0) \bJ'(\bphi_0) \right| \,.
    \end{equation} 
    \noindent Using (\ref{eq:sub_preturb}) and derivations that are elaborated in appendix \ref{appendix}  we simplify expressions for the gradient to
    \begin{equation}
        \label{eq:gradient_hessian2}
        \bJ'(\bphi_0) \approx 2\text{Re} \left \{\text{diag}[\bA^{\dagger}(\bU_s\bW\bLambda_s^{-1}\bU_s^H\bA\text{diag}(\be)\bD^H)\bP_A^{\perp}\bD]\right\}\,.
    \end{equation} 
    The expressions for the gradient and Hessian can be further simplified for special choices of the weighting matrix $\bW$. We will consider cases when there is no weighting $\bW = \bI$ and when $\bW = \bLambda_s + \sigma_n^2\bI$.
    
    \textit{(i)} Lets assume that $\bW = \bI$, then, by expressing covariance matrix of MPCs amplitudes as $\bR_{\tilde{\balpha}}^{-1} = \bA^H\bU_s\bLambda_s^{-1}\bU_s^H\bA$ \cite{stoica1989music}, we can write the gradient as  
    \begin{equation}
        \label{eq:gradient3}
        \bJ'(\bphi_0) \approx 2\text{Re} \left \{\text{diag}[(\bA^H\bA)^{-1}\text{diag}(\bsigma_{\tilde{\alpha}}^{-1}\odot\be)\bD^H\bP_A^{\perp}\bD]\right\}\,.
    \end{equation} 
    Likewise, we can reduce the expression for the Hessian to
    \begin{equation}
        \label{eq:hessian3}
        \overline{\bJ}''(\bphi_0) = - 2\text{Re} \left\{\left[\bD^H\bP_A^{\perp}\bD\right] \odot\left[(\bA^H\bA)^{-1}\right]^T\right\}\,.
    \end{equation} 

    \textit{(ii)} Similarly, when $\bW = \bLambda_s + \sigma_n^2\bI$ the gradient can be reduced to
     \begin{align}
        \label{eq:gradient4}
        \nonumber
        & \bJ'(\bphi_0) \\
        &\approx  2\text{Re} \left \{\text{diag}[(\bI + \sigma_n^2(\bA^H\bA)^{-1}\text{diag}(\bsigma_{\tilde{\alpha}}^{-1}))\text{diag}(\be)\bD^H\bP_A^{\perp}\bD]\right\}\,.
    \end{align} 
    \noindent  We see from (\ref{eq:error}) and (\ref{eq:gradient3}) that the bias introduced by unresolved MPCs to the delay estimate of the MPC of interest is proportional to the product of the power of interfering MPCs and their delay difference compared to the desired MPC. At the same time, this bias is inversely proportional to the total power of all MPCs in the same cluster. From (\ref{eq:gradient4}) we can conclude that bias also depends on the choice of weighting matrix.
    
   \section{Numerical Experiments}
    Numerical experiments are conducted to verify the derived analytical expression for the bias introduced by unresolved MPCs. We consider that the receiver probes the channel and estimates the channel frequency response in $L=4$ bands, using a probing signal with $N=12$ subcarriers and a bandwidth of $B=12$ MHz. We assume that the channel is probed multiple times during channel coherence time and we set the number of collected snapshots to $32$. The central frequencies of the  band's are set to $\left\{ 10, 50, 80, 150\right\}$ MHz, respectively. To evaluate the performance of the delay estimation, we use the Root Mean Square Error (RMSE) of the LOS delay estimate. The RMSE for the biased estimation of phase shift introduced by MPCs delays is defined as \cite{kay1998estimation, van2004optimum}
    \begin{equation}
        RMSE(\hat{\bphi}) := \sqrt{\text{var}(\hat{\bphi}) + BIAS^2(\hat{\bphi})}\,,
    \end{equation}
    \noindent where $\text{var}(\hat{\bphi})$ is the variance of the estimator in case when all MPCs are resolved and bias is not present. The Cram\'er Rao Lower Bound (CRLB) derived in \cite{stoica1989music} sets the lower bound on the variance and we use it to incorporate finite sampling effects in $RMSE(\hat{\bphi})$.
    The average RMSE is computed using $10^3$ independent Monte-Carlo trials and compared with the derived expression for the bias.
    
\begin{figure}[t!]
    	\centering
    	\subfloat[]{%
    		\includegraphics[trim=1 1 0 1,clip, width=6.6cm]{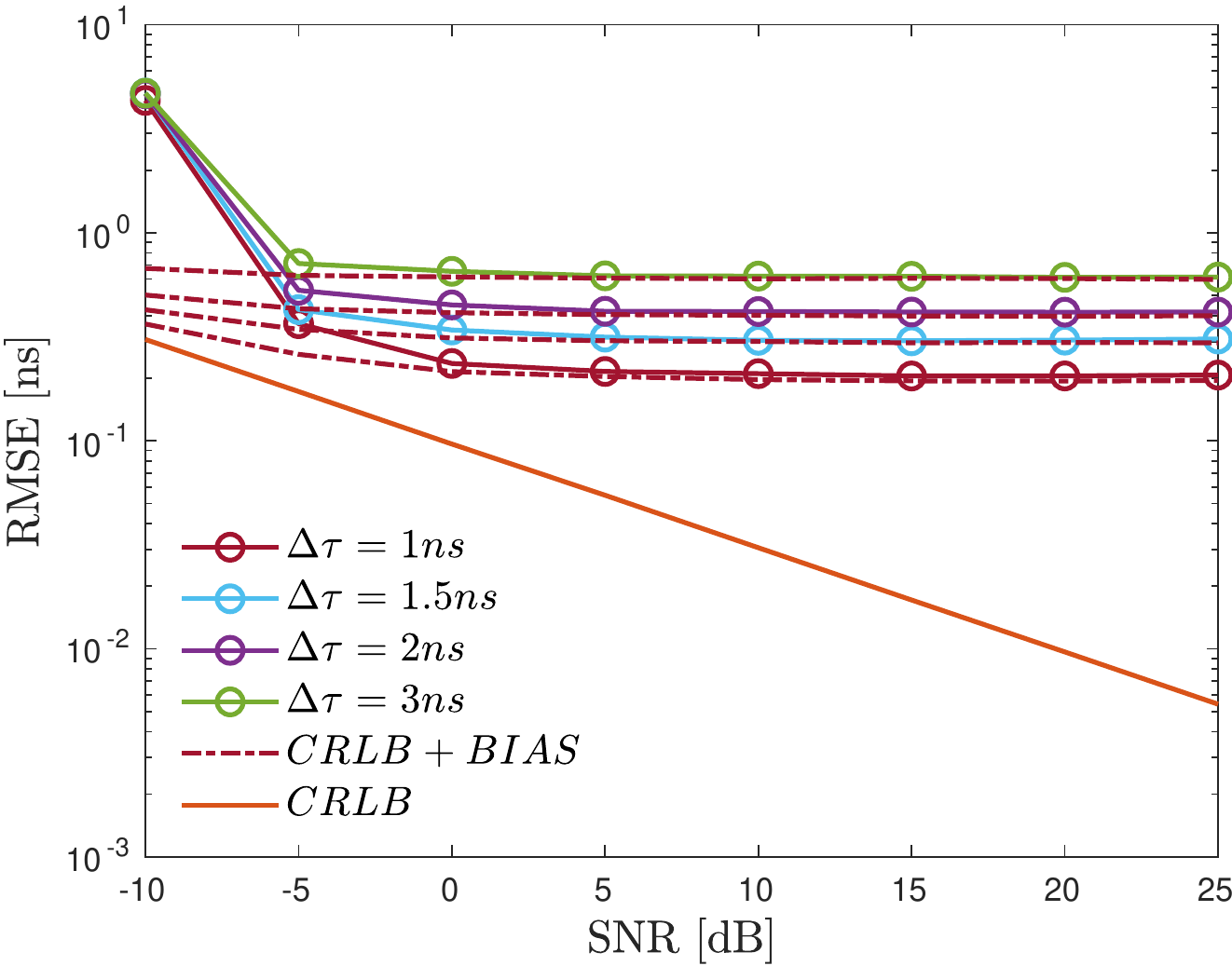}%
    		\label{fig:res:r1}%
    	}\qquad
    	\subfloat[]{%
    		\includegraphics[trim=0 1 1 2,clip,width=6.6cm]{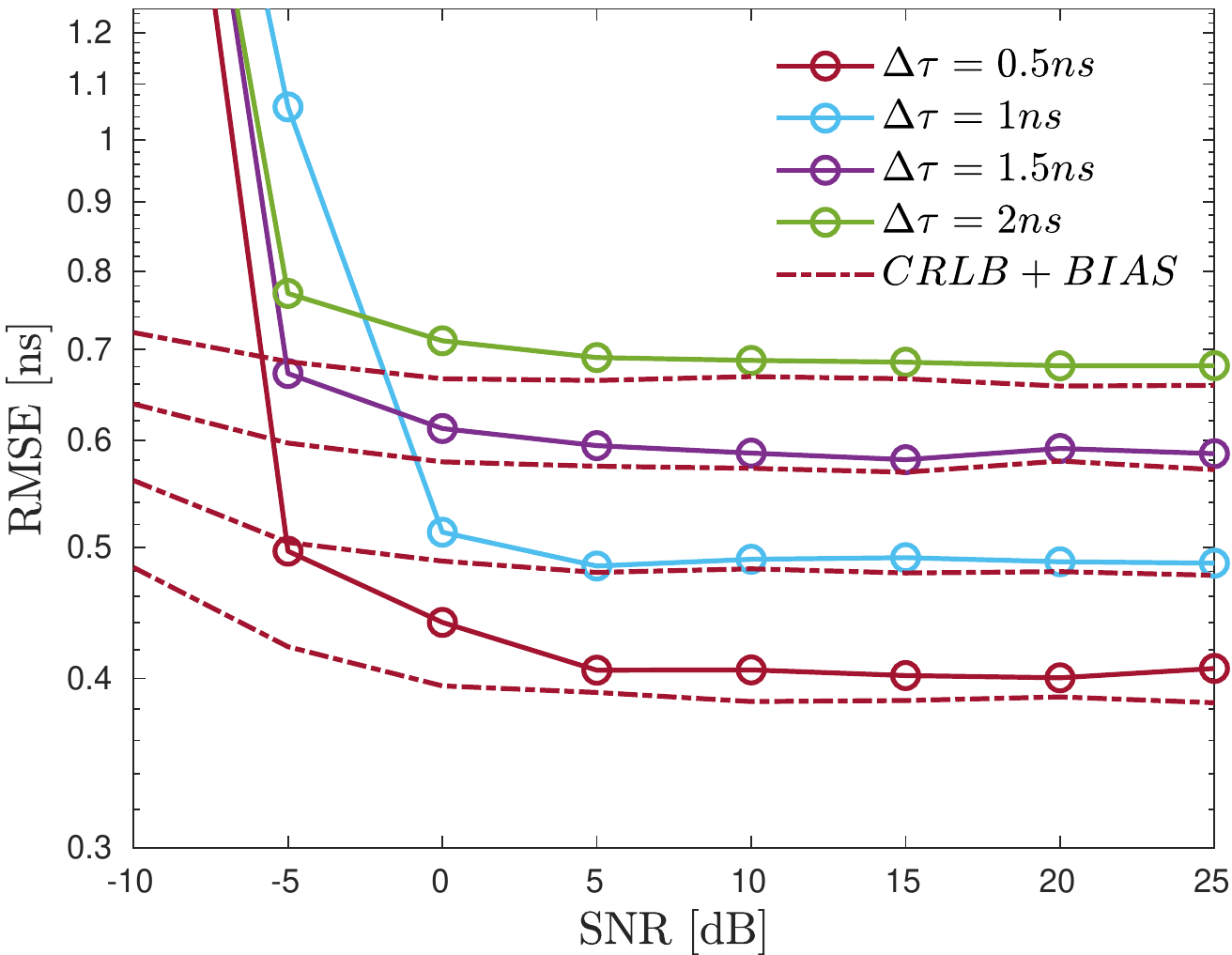}%
    		\label{fig:res:r2}%
    	}\qquad
        \subfloat[]{%
    		\includegraphics[trim=0 1 1 2,clip,width=6.6cm]{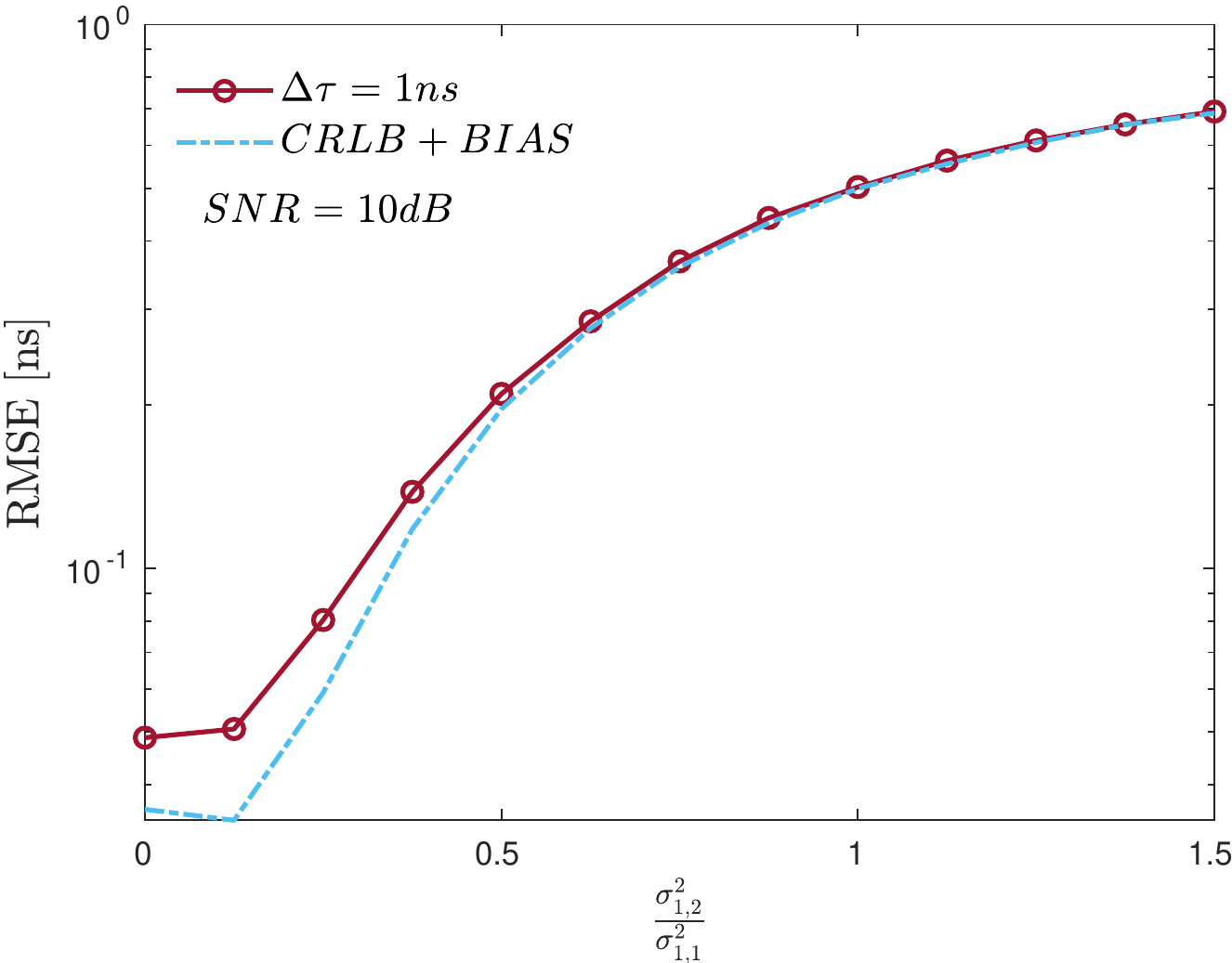}%
    		\label{fig:res:r3}%
    	}\qquad
    	\caption{Root Mean Square Error (RMSE) of the LOS delay estimate ($\tau_{1,1}$) for: (a) single unresolved MPC (b) two unresolved MPCs and (c) varying power of the single unresolved MPC .}
    	\label{fig:res}
    \end{figure}

    In the first simulation scenario, we consider three clusters of MPCs, i.e., $P=3$, where the clusters have $\left\{2, 3, 1 \right\}$ underlying multipath components with their powers set to $\left\{1, 0.5, 0.85, 0.55, 0.35, 0.55\right\}$, respectively. The delays of the LOS component and MPCs in the second and third cluster are kept fixed and set to $\left\{5, 33, 33.5, 34, 95\right\}$ ns, while the delay for the second MPC in the first cluster is changing during the trials and takes the values in $\left\{6, 6.5, 7, 8\right\}$ ns. The signal to noise ratio (SNR) is varied during the trails. From Fig.\ \ref{fig:res:r1}, it can be seen that the derived expression for the bias and CRLB set a tight bound on the expected RMSE of the delay estimation. However, in the low-SNR regime, the finite sampling and noise effects are dominant compared to errors introduced by unresolved MPCs. Furthermore, it can be observed that when the delay between the LOS and the interfering component increases, the bias on the delay estimate of the LOS component is also increasing. 
    
     In the second simulation scenario, we assume that in the first cluster, there are three underlying MPCs with their power set to $\left\{1, 0.5, 0.37 \right\}$, respectively. The delays of LOS component and MPCs in the second and third cluster are kept fixed and are the same as in the previous scenario, while the delay of third MPC in the first cluster is set to $8$ ns. Similar as at an earlier scenario the delay of the second MPC in the first cluster is changing during trials and takes the values in $\left\{5.5, 6, 6.5, 7\right\}$ ns. In Fig.\ \ref{fig:res:r2}, we observe that as in the previous scenario, the derived expression for the bias well describes the algorithm's performance. 
     
     In the third simulation, we consider the scenario where the power of the second MPC in the first cluster is changing relative to the LOS component's power. The SNR is set to $10$ dB, and the second MPC delay is set to $10$ ns, and they are kept fixed during simulations, while other parameters are the same as in the first scenario. As expected, from Fig.\ \ref{fig:res:r3}, we observe that when the power of the second MPC increases, the bias of the delay estimates also increases. We further notice that in the regime where the power of the second MPC is small, there is a gap between the derived bound and RMSE of the simulations. The reason for this is that in this case, finite sampling and noise errors dominate the bias introduced by unresolved MPC. 
     
    \section{Conclusions}
    In this paper, we analyze the effects of unresolved MPCs on the performance of delay estimation using a multiband subspace fitting algorithm. We treat this problem as a model error that leads to perturbed subspace estimation. We approximate subspace perturbations using the first-order Taylor expansion and derive the expression for the bias of the delay estimate of the LOS component. The resulting expression shows that this bias depends on the choice of the weighting matrix, the powers and the relative delays of the unresolved MPCs in the first cluster, compared to the LOS component.
  
    \begin{appendices}
    \section{Appendix}
    \label{appendix}
    The expression for the gradient $\bJ'(\bphi)$ (\ref{eq:gradient_hessian}) can be written as $\bJ'(\bphi) =  2\text{Re}\{\text{diag}[\bA^{\dagger}(\bU_s+\bU_e)\bW(\bU_s+\bU_e)^H\bP_A^{\perp}\bD]\}$. Assuming that second order perturbations are small  \cite{liu2008first} and using the property that in the noiseless case $\bP_{\bU_s}^{\perp} = \bP_A^{\perp}$ \cite{stoica1989music}, we can write the gradient as
    \begin{equation}
    \label{eq:grad_ext}
    \bJ'(\bphi) =  2\text{Re}\{\text{diag}[\bA^{\dagger}(\bU_s\bW\bU_e^H+\bU_e\bW\bU_s^H)\bP_A^{\perp}\bD]\}\,.
    \end{equation}
    Substituting (\ref{eq:sub_preturb}) in terms related to perturbations we can write $\bU_s\bW\bU_e^H = \sum_{p=1,i \neq p}^{P} [\bW]_{p,p} \bu_{s,p} \bu_{e,p}^H$ and $\bU_e\bW\bU_s^H = \sum_{p=1,i \neq p}^{P} [\bW]_{p,p} \bu_{e,p} \bu_{s,p}^H$, where $[]_{p,p}$ selects the $p$th entry in $p$th row of a matrix. Now, expanding $\bu_{e,p}$ and assuming that the contribution of the signal subspace to the perturbation is small \cite{xu2002perturbation}, we can write 
    \begin{align}
    \label{eq:app}
    \nonumber
        &\bU_s\bW\bU_e^H+\bU_e\bW\bU_s^H \\
        &= \sum_{p=1,i \neq p}^{P} [\bW]_{p,p} \sum_{m=1}^{LM-P} \beta_{i,m} (\bu_{s,p} \bu_{n,m}^H +  \bu_{n,m} \bu_{s,p}^H)\,.
    \end{align}
    Substituting the expression for $\beta_{i,m}$ in (\ref{eq:grad_ext}) and noticing that the second term in the sum (\ref{eq:app}) is equal to zero after multiplication by $\bP_A^{\perp}$, we can write 
    \begin{equation}
        \label{eq:grad_ext2}
        \bA^{\dagger}(\bU_s\bW\bU_e^H+\bU_e\bW\bU_s^H)\bP_A^{\perp}\bD = \bA^{\dagger}(\bU_s\bW\bLambda_s^{-1}\bU_s^H\bE)\bP_A^{\perp}\bD\,.
    \end{equation}
    \noindent Likewise, after substitution of $\bE$ in (\ref{eq:grad_ext2}), its second term is equal to zero due to the multiplication by $\bP_A^{\perp}$. Finally, substituting this in (\ref{eq:grad_ext}) results in (\ref{eq:gradient_hessian2}). 
    \end{appendices}

\bibliographystyle{IEEEtran}
\bibliography{paper}

\end{document}